\documentclass[aps,twocolumn,superscriptaddress,showpacs,amsmath]{revtex4-1}
\usepackage{graphicx}
\usepackage[colorlinks=true,citecolor=blue]{hyperref}

\let\Im\relax
\DeclareMathOperator{\Im}{Im}
\let\Re\relax
\DeclareMathOperator{\Re}{Re}
\renewcommand{\vec}[1]{\mathbf{#1}}

\newcommand{\ket}[1]{|#1\rangle}

\begin{document}

\title{The Majorana STM as a perfect detector of odd-frequency superconductivity}

\author{Oleksiy Kashuba}
\affiliation{Institute for Theoretical Physics and Astrophysics, University of W\"urzburg, D-97074 W\"urzburg, Germany}
\email{okashuba@physik.uni-wuerzburg.de}
\author{Bj\"orn Sothmann}
\affiliation{Theoretische Physik, Universit\"at Duisburg-Essen, D-47048 Duisburg, Germany}
\author{Pablo Burset}
\affiliation{Institute for Theoretical Physics and Astrophysics, University of W\"urzburg, D-97074 W\"urzburg, Germany}
\author{Bj\"orn Trauzettel}
\affiliation{Institute for Theoretical Physics and Astrophysics, University of W\"urzburg, D-97074 W\"urzburg, Germany}

\begin{abstract}
We propose a novel scanning tunneling microscope (STM) device in which the tunneling tip is formed by a Majorana bound state (MBS).
This peculiar bound state exists at the boundary of a one-dimensional topological superconductor.
Since the MBS has to be effectively spinless and local, we argue that it is the smallest unit that shows itself odd-frequency superconducting pairing.
Odd-frequency superconductivity is characterized by an anomalous Green function  which is an odd function of the time arguments of the two electrons forming the Cooper pair.
Interestingly, our Majorana STM can be used as the perfect detector of odd-frequency superconductivity.
The reason is that a supercurrent between the Majorana STM and any other superconductor can only flow if the latter system exhibits itself odd-frequency pairing.
To illustrate our general idea, we consider the tunneling problem of the Majorana STM coupled to a quantum dot in vicinity to a conventional superconductor.
In such a (superconducting) quantum dot, the effective pairing can be tuned from even- to odd-frequency behavior if an external magnetic field is applied to it.
\end{abstract}


\maketitle

The phenomenon of superconductivity (SC) comes in different facets.
Conventional SC, having granted us with a number of exciting micro- and macroscopic effects, is only a share of the whole zoo of superconducting phenomena.
In recent years, unconventional superconducting pairing~\cite{MineevSamokhin1999,Sigrist1991} has been proposed to exist in various forms, for instance, as the pairing mechanism for high-$T_{c}$ SC~\cite{Lee2006}, $p$-wave SC~\cite{ReadGreen2000}, topological SC with Majorana bound states as part of it~\cite{Kitaev2001,Ivanov2001}, and odd-frequency SC~\cite{Berezinskii1974,Balatsky1992,Bergeret2005,Tanaka2012,Eschrig2015}.
In this Letter, we will combine the latter two forms of unconventional SC to propose a new device -- the Majorana scanning tunneling microscope (STM).

\begin{figure}
\centering
\includegraphics[scale=1]{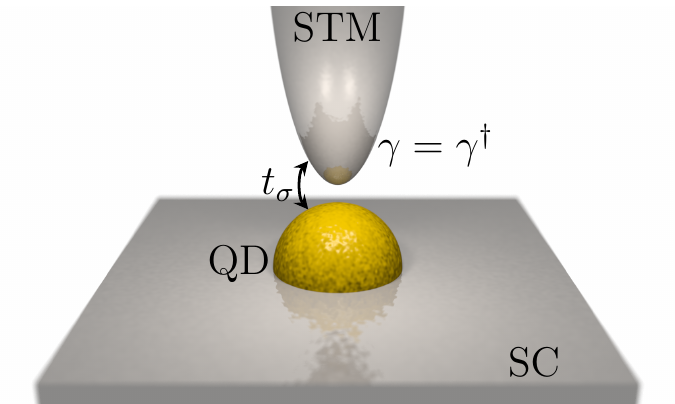}
\caption{The Majorana STM.
The scanning tip contains the Majorana bound state $\gamma$, which is used to probe -- via tunneling coupling $\textrm{t}_{\sigma}$ -- either an unknown superconductor with odd-frequency pairing or a quantum dot, which can realize odd-frequency pairing by proximity to an $s$-wave superconductor and an external magnetic field.}
\label{fig:STM}
\end{figure}

Let us start with describing the different ingredients of the Majorana STM, see Fig.~\ref{fig:STM} for a schematic.
Most importantly, we need a Majorana bound state (MBS), which has been predicted to exist at the boundary of a one-dimensional (1D) topological superconductor~\cite{Kitaev2001}.
A MBS can be induced into a spin-orbit coupled nanowire under the combined influence of conventional $s$-wave pairing and an external magnetic field~\cite{Lutchyn2010,Oreg2010}.
Recent experiments on the basis of nanowires and magnetic adatoms on $s$-wave superconductors have, indeed, shown some evidence that these exotic bound states which constitute their own ``antiparticles'' do exist in nature~\cite{Mourik2012,Das2012,Yazdani2014,Marcus2016}.
A MBS should form the tip of our STM, which can, for instance, be achieved by using a corresponding nanowire setup or, likewise, by any other realization of a 1D topological superconductor.
Now, the interesting question comes up how this device relates to odd-frequency pairing.

Odd-frequency SC is defined on the basis of the anomalous Green function that describes the superconducting pairing, {\it cf.}\ Eq.~(\ref{Fc}) below.
This Green function contains two annihilation operators corresponding to the particles that form the Cooper pair.
Due to the Pauli principle the Green function has to be odd under the exchange of these operators.
In the case of equal-time pairing, this oddness implies that singlet pairing has to be even in space coordinates and triplet pairing odd.
Interestingly, Berezinskii realized already in 1974~\cite{Berezinskii1974} that the symmetry of the pairing amplitude (proportional to the anomalous Green function) becomes richer if pairing at different times is allowed.
This was the birth of odd-frequency superconductivity where the oddness in time is transferred to the frequency domain by Fourier transformation.
Then, pairing mechanisms that are odd in frequency, triplet in spin space, and even in spatial parity (OTE)~\cite{Tanaka2012} are allowed by symmetry.
Exciting new physics is attributed to OTE pairing, e.g., related to a long-range proximity effect in hybrid Josephson junctions based on ferromagnetism and superconductivity~\cite{kadigrobov_quantum_2001,Bergeret2001}, cross-correlations between the end states in a topological wire~\cite{Huang2015}, the interplay of superconductivity and magnetism in double quantum dots~\cite{Sothmann2014} or its connection to crossed Andreev reflection at the helical edge of a 2D topological insulator~\cite{Crepin2015}.

Remarkably, a single MBS is the prime example for OTE superconductivity.
This somewhat surprising statement can be understood by very simple means: The annihilation operator $\gamma$ of the MBS is Hermitian, i.e.\ $\gamma=\gamma^\dagger$.
Thus, in the case of a MBS, the normal and the anomalous Green functions coincide, see Eq.~\eqref{eq:DGF} below.
Moreover, a single MBS has no additional quantum numbers, like spin, momentum, etc., i.e.\ it corresponds to a spinless, local object.
Thus, the time-ordered Majorana correlator $\langle \mathsf{T}\gamma(t)\gamma(0)\rangle$, where the symbol $\mathsf{T}$ denotes the time ordering, has to be antisymmetric in $t$ because of the Pauli principle.
This property is, in fact, in one-to-one correspondence to the emergence of odd-frequency superconductivity.

Therefore, it is natural to use this property of a MBS as building block for the Majorana STM.
If the tip of this device is formed by the MBS then a supercurrent from this tip can only flow into any other superconductor if and only if this superconductor also exhibits (at least partly) odd-frequency triplet SC.
If not the corresponding supercurrent completely vanishes for symmetry reasons.
It should be mentioned that it is rather difficult to detect odd-frequency SC.
Our novel idea constitutes a qualitative way of achieving this challenging task.
In the following, we will first describe our proposal based on general grounds and, subsequently, apply it to a concrete example.

\paragraph{Symmetry considerations.--} Odd-frequency SC can be best understood on the basis of the symmetry properties of the Green function that describes the anomalous (causal) correlation function~\footnote{%
The symmetries of the anomalous correlators and the manifestation of the fermionic anticommutativity are analyzed in the Appendix~\ref{apx:correlators}.
}\edef\notea{Note\thefootnote}%
, i.e.
\begin{equation} \label{Fc}
F^{c}_{\alpha\beta}(t) \!=\!
-i\langle \mathsf{T}\psi_{\alpha}(t)\psi_{\beta}(0)\rangle \!=\!
\tfrac{1}{2} (F^{K}_{\alpha\beta} + F^{R}_{\alpha\beta} + F^{A}_{\alpha\beta})(t).
\end{equation}
Here, $\psi_{\alpha}(t)$ is an annihilation operator for the electron in state $\alpha$ (encoding orbital and/or spin degrees of freedom) at time $t$; angle brackets denote the averaging over the ground state; and $F^{R/A/K}_{\alpha\beta}(t)$ are the retarded/advanced/Keldysh components~\cite{Keldysh1965} of the anomalous correlation functions specified below in Eq.~\eqref{eq:DGF}.
Due to the Pauli principle, the time-ordered Green function fulfills the following symmetry condition: $F^{c}_{\beta\alpha}(-t)=-F^{c}_{\alpha\beta}(t)$.
When we calculate transport properties below, not the time-dependent correlation functions $F^{R/A/K}_{\alpha\beta}(t)$ matter but instead their Fourier transforms $F^{R/A/K}_{\alpha\beta}(\omega)$.
Therefore, it is important to state how the retarded, the advanced, and the Keldysh Green functions behave under a sign change of $\omega$.
This behavior is summarized in Table~\ref{tab:symm}.
In this table, we do not only refer to the symmetry properties of the anomalous Green functions $F^{R/A/K}_{\alpha\beta}(\omega)$ (relevant for the superconducting properties of the system) but, for completeness, also to the normal Green functions $G^{R/A/K}_{\alpha\beta}(\omega)$.

In thermal equilibrium (at temperature $T$), the Keldysh Green function can be expressed as a simple function of retarded and advanced Green function via
\begin{equation}
X^{K}(\omega)=\tanh\frac{\omega}{2T}\left[X^{R}(\omega)-X^{A}(\omega)\right],
\label{eq:equilibrium}
\end{equation}
where $X$ could be the normal ($G$) or the anomalous ($F$) Green function.
Evidently, {\it cf.}\ Table~\ref{tab:symm}, some linear combinations of retarded, advanced, and Keldysh Green function are even with respect to frequency $\omega$ and others are odd.
Therefore, we need to carefully address their influence on the current that will flow through the Majorana STM to fully understand why this device functions as a perfect detector for odd-frequency SC.
We now develop a general microscopic model for the Majorana STM.
Specifically, we derive a formula for the Josephson supercurrent between the superconducting STM tip and an unknown SC as substrate.


\begin{table}
\begin{tabular}{rc|ccc}
 & & $R+A$ & $R-A$ & $K$ \\
 \hline
normal		& $G_{\alpha\beta}+G_{\beta\alpha}$\,	& $\Re$				& $\Im$				& $\Im$			\\
anomalous	& $F_{\alpha\beta}+F_{\beta\alpha}$\,	& odd				& even				& odd			\\
Majorana	& $D$\,									& \,\,$\Re$, odd\,	& \,$\Im$, even\,	& \,$\Im$, odd\,
\end{tabular}
\caption{Symmetry properties of the symmetrized (with respect to $\alpha,\beta$ space) Green functions: $\Re$/$\Im$ denotes whether the functions are real or imaginary; even/odd means whether the functions are even/odd in $\omega$.
$R \pm A$ should be understood as the corresponding linear combination of retarded and advanced Green function.}
\label{tab:symm}
\end{table}


\paragraph{Majorana STM.--} The coupling between the Majorana state $\gamma$ and another system can be described by the tunneling Hamiltonian~\cite{Leijnse2011}
\begin{equation}
H_{t} = \gamma \sum_{\alpha}\textrm{t}_{\alpha} \left(\psi_{\alpha} - \psi_{\alpha}^{\dagger} \right) = \sum_{\alpha}\textrm{t}_{\alpha}\left(\gamma \psi_{\alpha} + \psi_{\alpha}^{\dagger}\gamma \right),
\label{eq:Ht}
\end{equation}
where $\alpha$ denotes the different quantum numbers (e.g.\ spin, momentum, etc.), $\psi_{\alpha}$ is the annihilation operator of the scanned substrate, and $\textrm{t}_{\alpha}$ is the tunneling amplitude~\footnote{%
%
All gauge transformations, as well as the derivation of Eq.~\eqref{eq:IZ}, are addressed in the Appendix~\ref{apx:current}.
}\edef\noteb{Note\thefootnote}%
.
Then, the current operator can be written as
\begin{equation}
\hat{I} = e\dot{N} = i[H_{t},N]_{-} = i\frac{e}{\hbar} \,\gamma \sum_{\alpha}\textrm{t}_{\alpha}\left(\psi_{\alpha} + \psi_{\alpha}^{\dagger}\right),
\end{equation}
where $N$ is the number of electrons in the studied superconductor.
Hence, the average current is given by
\begin{equation}
I = \frac{e}{\hbar}\sum_{\alpha}\textrm{t}_{\alpha}\Re \int \frac{d\omega}{2\pi} W_{\alpha}^{K}(\omega),
\label{eq:I}
\end{equation}
where the integrand is the Fourier transformed Keldysh component
of the cross correlator $W_{\alpha}^{K}(t) = -i\langle [\psi_{\alpha}(t),\,\gamma(0)]_{-} \rangle$, which can be calculated exactly by means of the Dyson formula
\begin{multline}
W^{K}_{\alpha}(\omega) = \sum_{\beta}\textrm{t}_{\beta}
\left\{
\left[ G^{(0)R}_{\alpha\beta}(\omega) - F^{(0)R}_{\alpha\beta}(\omega) \right] D^{K}(\omega) +\right.\\\left.+
\left[ G^{(0)K}_{\alpha\beta}(\omega) - F^{(0)K}_{\alpha\beta}(\omega) \right] D^{A}(\omega)
\right\}.
\label{eq:WDyson}
\end{multline}
In this expression, the functions of $\omega$ are the Fourier transforms of the retarded/advanced/Keldysh components of the Majorana Green function $D$, and normal (anomalous) Green function of the lead $G$ ($F$), which are defined as
\begin{equation}
\begin{split}
D^{R/A}(t) &=\mp i \langle \{\gamma(t),\,\gamma(0)\}_{+} \rangle \theta(\pm t),
\\
D^{K}(t)   &=  - i \langle [\gamma(t),\,\gamma(0)]_{-}\rangle,
\\
G_{\alpha\beta}^{R/A}(t) &= \mp i\langle \{\psi_{\alpha}(t),\,\psi_{\beta}^{\dagger}(0)\}_{+} \rangle \theta(\pm t),
\\
G_{\alpha\beta}^{K}(t) &= -i\langle [\psi_{\alpha}(t),\,\psi_{\beta}^{\dagger}(0)]_{-} \rangle,
\\
F_{\alpha\beta}^{R/A}(t) &= \mp i\langle \{\psi_{\alpha}(t),\,\psi_{\beta}(0)\}_{+} \rangle \theta(\pm t),
\\
F_{\alpha\beta}^{K}(t) &= -i\langle [\psi_{\alpha}(t),\,\psi_{\beta}(0)]_{-} \rangle.
\end{split}
\label{eq:DGF}
\end{equation}
The superscript $\vphantom{G}^{(0)}$ in Eq.~\eqref{eq:WDyson} denotes that the Green functions are bare with respect to the tunneling Hamiltonian $H_{t}$ in Eq.~\eqref{eq:Ht}.
Substituting the cross correlator from Eq.~\eqref{eq:WDyson} into the expression for the current in Eq.~\eqref{eq:I} and removing all vanishing terms due to the mismatching symmetries with respect to $\omega$, we obtain
\begin{multline*}
I = \frac{e}{2\hbar}\sum_{\alpha\beta}\textrm{t}_{\alpha}\textrm{t}_{\beta} \int \frac{d\omega}{2\pi}
\Bigl\{
\\-
\Im\left[G^{(0)R}_{\alpha\beta}(\omega) - G^{(0)A}_{\alpha\beta}(\omega)\right] \Im D^{K}(\omega)
+\\+
\Im G^{(0)K}_{\alpha\beta}(\omega)\Im\left[D^{R}(\omega)-D^{A}(\omega)\right]
+\\+
\Im\left[F^{(0)R}_{\alpha\beta}(\omega) + F^{(0)A}_{\alpha\beta}(\omega) \right] \Im D^{K}(\omega)
-\\-
\Re F^{(0)K}_{\alpha\beta}(\omega)\Re\left[D^{R}(\omega)+D^{A}(\omega)\right]
\Bigr\}.
\end{multline*}
Note that this current takes into account both normal current and supercurrent contributions.
We are interested in the supercurrent only which can flow when both tip and substrate are in mutual thermal equilibrium.
This assumption reduces the number of independent Keldysh Green functions according to Eq.~\eqref{eq:equilibrium}.
Then, the terms proportional to the normal Green functions cancel out and we are left with the expression for the supercurrent
\begin{multline} \label{I_SC}
I = \frac{e}{2\hbar}\sum_{\alpha\beta}\textrm{t}_{\alpha}\textrm{t}_{\beta} \int \frac{d\omega}{2\pi}\tanh\frac{\omega}{2T}
\times\\\times
\left\{
\Im\left[F^{(0)R}_{\alpha\beta}(\omega) \!+\! F^{(0)A}_{\alpha\beta}(\omega) \right]
\!
\Im\left[ D^{R}(\omega) \!-\! D^{A}(\omega) \right]
\!-\!\right.\\\left.\!-\!
\Re\left[F^{(0)R}_{\alpha\beta}(\omega) \!-\! F^{(0)A}_{\alpha\beta}(\omega)\right]
\!
\Re\left[D^{R}(\omega) \!+\! D^{A}(\omega)\right]
\right\}.
\end{multline}
If we compare the integrand of the latter equation with the symmetry properties of the Green functions in Table~\ref{tab:symm}, we evidently see that only odd contributions to the anomalous correlation functions (that describe the scanned substrate) can contribute to the supercurrent~\cite{\notea}.
This constitutes the main result of our Letter.
In order to express the current by means of correlators of the investigated superconductor only, we need to know the full Majorana Green function $D$, the self-energy of which can be written as
\begin{multline}
\Sigma^{R/A}(\omega) = \sum_{\alpha\beta} \textrm{t}_{\alpha}\textrm{t}_{\beta} \Bigl\{
  G^{(0)R/A}_{\alpha\beta}(\omega)
- G^{(0)A/R}_{\alpha\beta}(-\omega)
-\\
- F^{(0)R/A}_{\alpha\beta}(\omega)
-\left[F^{(0)A/R}_{\alpha\beta}(\omega)\right]^{*}
\Bigr\} \; .
\end{multline}
Note that the self-energy obeys the same symmetry relations as the Majorana Green function $D$, i.e.\ $\Sigma^{R}(\omega)=[\Sigma^{A}(\omega)]^{*}=-\Sigma^{A}(-\omega)$.
The bare Majorana Green function is $D^{(0)R/A}=2(\omega\pm i0)^{-1}$, so the full Majorana Green function becomes $D^{R/A}=2(\omega - 2\Sigma^{R/A}(\omega))^{-1}$. These expressions can be plugged into Eq.~(\ref{I_SC}) to further evaluate the supercurrent for a particular substrate under consideration.
We now illustrate our general result on the basis of a concrete example where the amount of odd-frequency SC can be easily tuned.


\paragraph{Superconducting quantum dot substrate.--} Let us consider a single-level quantum dot with Coulomb energy $U$ subject to an external magnetic field $\vec B=(B_{\perp}\cos\theta,B_{\perp}\sin\theta,B_{z})$ pointing in an arbitrary direction with respect to the spin quantization axis that is effectively defined by the MBS at the tip of the STM.
The magnetic field acts only on the spin degree of freedom $\hat{\vec S}=\frac{1}{2}\sum_{\sigma\sigma'}c^{\dagger}_{\sigma}\boldsymbol{\sigma}_{\sigma\sigma'}c_{\sigma'}$ of the dot, where $\boldsymbol{\sigma}$ is the vector of Pauli matrices.
The dot level with energy $\varepsilon$ is coupled via the tunnel coupling $\Gamma_{\Delta}$ to a conventional $s$-wave superconductor
%
%
with order parameter $\Delta e^{i\phi}$. In the following, we will focus on subgap transport where quasiparticle contributions are exponentially suppressed in $\Delta/T$. This allows us to integrate out the superconducting degrees of freedom, leave aside sophisticated Kondo physics~\cite{Lutchyn2014}, and leads to an effective dot Hamiltonian~\cite{rozhkov_interacting-impurity_2000,sothmann_probing_2010}
\begin{equation}
H_\text{dot} \!=\! \sum_\sigma \varepsilon c_{\sigma}^\dagger c_{\sigma} \!+\!
\mathbf{B}\cdot \hat{\mathbf{S}} \!+\!
U n_\uparrow n_\downarrow \!-\!
\frac{\Gamma_{\Delta}e^{i\phi}}{2}c_\uparrow^\dagger c_\downarrow^\dagger+\text{H.c.}
\label{eq:Hdot}
\end{equation}
The eigenstates of the isolated quantum dot-superconductor system are given by states of a single occupied dot with spin parallel $\ket{\uparrow_{\mathbf{B}}}$ and antiparallel $\ket{\downarrow_{\mathbf{B}}}$ to the magnetic field with energies $E_{\uparrow/\downarrow}=\varepsilon\pm B/2$.
Furthermore, there exist the mixtures of empty $\ket{0}$ and fully occupied $\ket{\uparrow\downarrow}$ dot states
\begin{equation*}
|\pm\rangle=\frac{1}{\sqrt{2}}
\left(\sqrt{1\mp\frac{\delta}{2\varepsilon_\text{A}}}|0\rangle \mp \sqrt{1\pm\frac{\delta}{2\varepsilon_\text{A}}}|\uparrow\downarrow\rangle\right)
\end{equation*}
with energies $E_\pm=\frac{\delta}{2}\pm\varepsilon_\text{A}$, where $\varepsilon_\text{A}=\frac{1}{2}\sqrt{\delta^2+\Gamma_{\Delta}^2}$ and $\delta=2\varepsilon+U$.


In order to characterize the superconducting correlations induced on the quantum dot, we consider the time-ordered anomalous Green functions
$F^{c}_{\sigma\sigma'}(t)=\langle \mathsf{T} c_{\sigma'}(t)c_\sigma(0) \rangle$ which can be written in terms of the density matrix elements of the quantum dot $\langle\alpha|\rho|\beta\rangle=P_\alpha\delta_{\alpha\beta}$, where $|\alpha\rangle$ are the eigenstates of the Hamiltonian~\eqref{eq:Hdot} given above, $P_\alpha=Z^{-1}e^{-E_{\alpha}/T}$, and $Z=\sum_{\alpha}e^{-E_{\alpha}/T}$. As a result, we arrive at
\begin{multline}
F^{c}_{\sigma'\sigma}(t) = \sum_{\alpha\beta}\int\frac{d\omega}{2\pi} e^{i\omega t}P_{\alpha}\Biggl(
\frac{ \langle\alpha|c_\sigma|\beta\rangle \langle\beta|c_{\sigma'}|\alpha\rangle
}{ \omega-E_\beta+E_\alpha+i0^+}
+\\+
\frac{ \langle\alpha|c_{\sigma'}|\beta\rangle \langle\beta|c_\sigma|\alpha\rangle
}{ \omega+E_\beta-E_\alpha-i0^+}\Biggr).
\end{multline}
Parametrizing the Green functions as $F^c_{\sigma\sigma'}(t)=\left\{i\left[F^c_s(t)+\vec F^c_t(t)\cdot\boldsymbol{\sigma}\right]\sigma_y\right\}_{\sigma\sigma'}$, we can define an effective order parameter for the singlet part which corresponds to the even-frequency component and is equal to
\begin{equation}
F^{c}_s(0)=\frac{i\pi\Gamma_{\Delta}}{2\varepsilon_\text{A}}(P_+-P_-).
\label{eq:Feven}
\end{equation}
To characterize the triplet part, we employ the time derivative of the Green function as an effective order parameter of the odd-frequency component~\cite{Balatsky1993,Schrieffer1994,Abrahams1995,Dahal2009} to obtain
\begin{equation}
\partial_{t}\vec F^{c}_t(0) = \pi\Gamma_{\Delta}\vec S-\frac{i}{2}\vec BF^c_s(0).
\label{eq:Fodd}
\end{equation}
Hence, these order parameters depend on the expectation value of the spin operator $\vec S= \sum_{\alpha}P_\alpha\langle\alpha|\hat{\vec S}|\alpha\rangle$ of the quantum dot and the magnetic field $\vec{B}$.

The coupling between the dot level and the MBS on the tip is given by the tunneling Hamiltonian~\eqref{eq:Ht} (with $\psi_{\alpha}=c_{\sigma}$ and $\textrm{t}_\alpha = \textrm{t}_\sigma$).
In the following, we represent the MBS $\gamma$ by a conventional spinless (nonlocal) fermion $f$ as $\gamma=f+f^\dagger$.
(This representation implies that there is a second MBS on the STM far away from the tunneling tip, which naturally happens in any realization of a 1D topological superconductor.)
Then, the full Hamiltonian decomposes into two blocks corresponding to even and odd parity of the total quantum dot/nonlocal fermion system.
As both blocks are equivalent to each other, we now focus on the odd parity sector.
It is spanned by the states $|\!\!\uparrow\downarrow,1\rangle$, $|\!\!\uparrow,0\rangle$, $|\!\!\downarrow,0\rangle$ and $|0,1\rangle$ where the first ket entry denotes the dot occupation while the second ket entry is the occupation of the nonlocal fermion described by the operator $f^\dagger f$.
Choosing the spin quantization axis such that $\textrm{t}_\uparrow=\textrm{t}$ is real and $\textrm{t}_\downarrow=0$, the Hamiltonian takes the form~\cite{\noteb}
\begin{equation}
H \!=\!
\begin{pmatrix}
	\delta & 0 & \textrm{t} & -\frac{\Gamma_{\Delta}}{2}e^{i\phi-i\theta} \\
	0 & \varepsilon+\frac{B_{z}}{2} & \frac{B_{\perp}}{2} & -\textrm{t} \\
	\textrm{t} & \frac{B_{\perp}}{2} & \varepsilon-\frac{B_{z}}{2} & 0 \\
	-\frac{\Gamma_{\Delta}}{2}e^{-i\phi+i\theta} & -\textrm{t} & 0 & 0
\end{pmatrix}.
\label{eq:Hdot+M}
\end{equation}
The eigenvalues of this Hamiltonian are the energies $E_{\alpha}(\phi-\theta)$ corresponding to many-body eigenstates $|\alpha\rangle$ which depend on the superconducting phase.
Then, the supercurrent can be calculated via the derivative of the free energy with respect to the phase~\cite{\noteb},
\begin{equation}
I(\phi)=\frac{2e}{\hbar}\partial_{\phi}\left(-T \log Z\right) = \frac{2e}{\hbar} \sum_{\alpha} P_{\alpha}\partial_{\phi}E_{\alpha}(\phi-\theta).
\label{eq:IZ}
\end{equation}
We find that a supercurrent can only flow if the direction of the external magnetic field is not collinear with the spin quantization axis of the MBS, i.e.\ $B_{\perp}\ne 0$.

\begin{figure}
\centering
\includegraphics[width=\columnwidth]{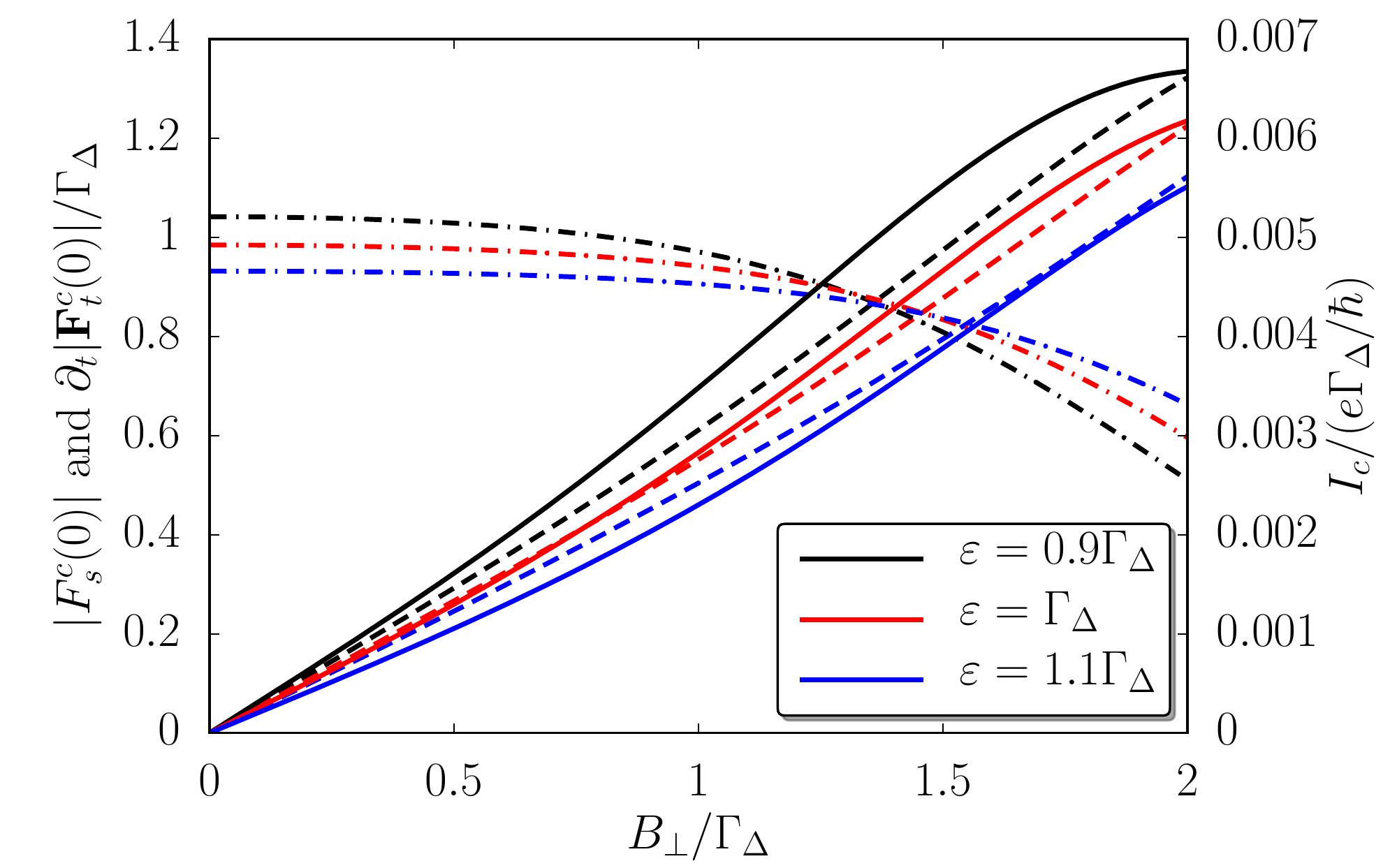}
\caption{
The critical current $I_c$ (solid lines) in comparison with the odd-frequency order parameter $\partial_t |\vec F^c_t(0)|$ (dashed lines) and the even-frequency order parameter $|F^c_s(0)$ (dashed-dotted lines) for $B_{z}=0$, $U=\Gamma_{\Delta}$, weak coupling $\textrm{t}=0.1\Gamma_{\Delta}$, and temperature $T=0.2\Gamma_{\Delta}$.
}
\label{fig:IvsF}
\end{figure}

The current is $2\pi$ periodic [with respect to $\phi$, as well as the Hamiltonian in Eq.~\eqref{eq:Hdot+M}] with a dominating first harmonics, so one can approximate $I(\phi) \approx I_{c}\sin(\phi-\theta)$. 
%
%
Interestingly, the influence of the angle $\theta$ on the current allows for a so-called $\phi_{0}$-junction~\cite{Buzdin2008} in our setup.
The dependence of $I_{c}$ on the system parameters is shown in Fig.~\ref{fig:IvsF}.
Evidently, an unambiguous correlation between the supercurrent and the odd-frequency pairing defined in Eq.~\eqref{eq:Fodd} can be observed.


\paragraph{Conclusions.--} We suggest a new device corresponding to a Majorana bound state at the tip of a scanning tunneling microscope, which we dub Majorana STM.
It is shown that a single Majorana bound state exhibits a pair amplitude that is an odd function of time.
This feature is decisive that the Majorana STM serves as an ideal detector for odd-frequency superconductivity.
If a supercurrent builds up between the Majorana STM and an unknown superconducting substrate then the latter superconductor has to experience odd-frequency pairing itself.
We illustrate this general result on the basis of a simple quantum dot model coupled to the Majorana STM.

\begin{acknowledgments}
Financial support by the DFG (SPP1666 and SFB1170 "ToCoTronics"), the Helmholtz Foundation (VITI), the ENB Graduate school on "Topological Insulators", and the Ministry of Innovation NRW is gratefully acknowledged.
\end{acknowledgments}

\bibliography{fosc}


\appendix

\onecolumngrid

\section{Correlators, Green functions, and their symmetries}
\label{apx:correlators}

In this section, we study the symmetries of the general correlator, apply the obtained result to the fermionic Green functions, and examine the manifestation of the Fermion anticommutativity in the different types of Green functions.

Let us consider two different objects described by the operators $A$ and $B$ in Heisenberg representation.
We can define three correlators that account for the different order of the operators with respect to the time coordinates, or for their hermitian conjugated versions.
Expressing the resulting Green functions on the Keldysh contour~\cite{Keldysh1965,RammerSmith1986}, we can write
\begin{equation}
\mathcal{G}_\mathsf{K}(t_{i},t_{j}') = -i\langle\mathrm{T}_\mathsf{K}A(t_{i})B(t_{j}')\rangle,
\quad
\widetilde{\mathcal{G}}_\mathsf{K}(t_{i},t_{j}') = -i\langle\mathrm{T}_\mathsf{K}B(t_{i})A(t_{j}')\rangle,
\quad
\overline{\mathcal{G}}_\mathsf{K}(t_{i},t_{j}') = - i\langle\mathrm{T}_\mathsf{K}B^{+}(t_{i})A^{+}(t_{j}')\rangle,
\label{eq:Gdef}
\end{equation}
where $\mathrm{T}_\mathsf{K}$ is the time-ordering operator on the Keldysh contour. 
Here, we follow a notation similar to Ref.~\onlinecite{RammerSmith1986}, where for the time coordinate $t_{i}$, the index $i=+(-)$ corresponds to the $c_{1}$ ($c_{2}$) contour that lies \emph{above} (\emph{below}) the time axis and is the \emph{first} (\emph{second}) part of the full Keldysh contour.
We thus express the Green functions in matrix form as
\begin{equation}
\mathcal{G}_\mathsf{K}(t_{i},t_{j}') = \bigl(\mathcal{G}_\mathsf{K}(t-t')\bigr)_{ij} = -i
\begin{pmatrix}
\langle\mathrm{T}A(t)B(t')\rangle & \mp \langle B(t')A(t)\rangle \\
\langle A(t)B(t')\rangle & \langle\widetilde{\mathrm{T}}A(t)B(t')\rangle
\end{pmatrix}
\equiv
\begin{pmatrix}
\mathcal{G}^{c} & \mathcal{G}^{<} \\
\mathcal{G}^{>} & \mathcal{G}^{ac}
\end{pmatrix}(t-t'),
\label{eq:GKdef}
\end{equation}
where $\mathrm{T}$ is the time-ordering operator, $\widetilde{\mathrm{T}}$ is the reverse time-ordering one, and the sign $-(+)$ corresponds to fermion (boson) operators. $\mathcal{G}^{c}$, $\mathcal{G}^{ac}$, $\mathcal{G}^{<}$, and $\mathcal{G}^{>}$ are causal, ``anti-causal'' (with reverse time ordering), ``greater'', and ``lesser'' Green functions. 
Note that
\begin{align*}
\langle\mathrm{T}A(t)B(t')\rangle &= \mp \langle\mathrm{T}B(t')A(t)\rangle, &
\langle\widetilde{\mathrm{T}}A(t)B(t')\rangle &= \mp \langle\widetilde{\mathrm{T}}B(t')A(t)\rangle, \\
\langle\mathrm{T}A(t)B(t')\rangle^{*} &= \langle\widetilde{\mathrm{T}}B^{+}(t')A^{+}(t)\rangle, &
\langle\widetilde{\mathrm{T}}A(t)B(t')\rangle^{*} &= \langle\mathrm{T}B^{+}(t')A^{+}(t)\rangle.
\end{align*}
These relations allow us to establish the connection between the different correlators defined in Eq.~(\ref{eq:Gdef}). Assuming that the Hamiltonian is time independent, the correlators depend on the time difference only, and are transformed into each other as
\begin{equation}
\widetilde{\mathcal{G}}_\mathsf{K}(t-t') = \mp \left(\mathcal{G}_\mathsf{K}(t'-t)\right)^\mathsf{T}
\qquad
\overline{\mathcal{G}}_\mathsf{K}(t-t') = - \tau^{1}\left(\mathcal{G}_\mathsf{K}(t'-t)\right)^{+}\tau^{1}
\end{equation}
where the $\vphantom{G}^{+}$ superscript denotes Hermitian conjugation in Keldysh matrix space and $\tau^{i}$ are the Pauli matrices acting on the same space.
The four Green functions defined in Eq.~\eqref{eq:GKdef} are, however, linearly dependent.
We can eliminate one component, if we rotate the basis of the Keldysh space~\cite{RammerSmith1986} as follows
\begin{gather}
\hat{\mathcal{G}}(t-t') = L\tau^{3}\mathcal{G}_\mathsf{K}(t-t')L^\mathsf{T}=
\begin{pmatrix}
\mathcal{G}^{R} & \mathcal{G}^{K} \\
0 & \mathcal{G}^{A}
\end{pmatrix}(t-t'),
\end{gather}
where $L=(1-i\tau^{2})/\sqrt{2}$ and $\mathcal{G}^{R,A,K}$ are retarded, advanced, and Keldysh Green functions, respectively.
Applying the same rotation to the other Green functions defined in Eq.~\eqref{eq:Gdef}, we get
\begin{equation}
\hat{\widetilde{\mathcal{G}}}(t) = \mp \tau^{1}\hat{\mathcal{G}}^\mathsf{T}(-t)\tau^{1},
\qquad
\hat{\overline{\mathcal{G}}}(t) = \tau^{2}\hat{\mathcal{G}}^{+}(-t)\tau^{2}.
\end{equation}
Performing the Fourier transform over the time variable, we get the relation between the different correlators in frequency representation, namely, 
\begin{align}
\widetilde{\mathcal{G}}^{R/A}(\omega) &= \mp \mathcal{G}^{A/R}(-\omega), &
\widetilde{\mathcal{G}}^{K}(\omega)   &= \mp \mathcal{G}^{K}(-\omega),
\label{eq:Gtilderel}
\\
\overline{\mathcal{G}}^{R/A}(\omega) &=   \bigl(\mathcal{G}^{A/R}(\omega)\bigr)^{*}, &
\overline{\mathcal{G}}^{K}(\omega)     &= - \bigl(\mathcal{G}^{K}(\omega)\bigr)^{*}.
\label{eq:Gbarrel}
\end{align}

\noindent
{\it The normal electron Green function} can be obtained from the definitions in Eq.~\eqref{eq:Gdef} by substitution of the generic operators $A=\psi_{\alpha}$ and $B=\psi_{\beta}^{+}$. It is thus defined as
\begin{equation}
{G_\mathsf{K}}_{\alpha\beta}(t_{i},t_{j}') = -i\langle\mathrm{T}_\mathsf{K}\psi_{\alpha}(t_{i})\psi_{\beta}^{+}(t_{j}')\rangle,
\end{equation}
where the indexes $\alpha$ and $\beta$ denote the full set of the electron quantum numbers, like spin, momentum, etc.
In that case, two of the correlators listed in Eq.~\eqref{eq:Gdef} are equivalent up to the exchange of the quantum numbers, namely,  ${\overline{G}_\mathsf{K}}_{\alpha\beta}(t_{i},t_{j}') = {G_\mathsf{K}}_{\beta\alpha}(t_{i},t_{j}')$.
Together with Eqs.~\eqref{eq:Gbarrel}, we obtain the symmetry of the normal electron Green function in frequency representation 
\begin{equation}
G^{R/A}_{\alpha\beta}(\omega) =   \bigl(G^{A/R}_{\beta\alpha}(\omega)\bigr)^{*},
\qquad
G^{K}_{\alpha\beta}(\omega)  = - \bigl(G^{K}_{\beta\alpha}(\omega)\bigr)^{*}.
\label{eq:Gsymm}
\end{equation}

\noindent
{\it The anomalous electron Green function} is defined via the substitution $A=\psi_{\alpha}$ and $B=\psi_{\beta}$, resulting in
\begin{equation}
{F_\mathsf{K}}_{\alpha\beta}(t_{i},t_{j}') = -i\langle\mathrm{T}_\mathsf{K}\psi_{\alpha}(t_{i})\psi_{\beta}(t_{j}')\rangle.
\end{equation}
Since the two generic operators are of the same type, we immediately find that $\widetilde{F}_{\mathsf{K}\,\alpha\beta}(t_{i},t_{j}') = {F_\mathsf{K}}_{\beta\alpha}(t_{i},t_{j}')$. 
As a result, the symmetry dependence with respect to the frequency of the anomalous Green function is
\begin{equation}
F^{R/A}_{\alpha\beta}(\omega) = - F^{A/R}_{\beta\alpha}(-\omega) 
\quad\text{and}\quad
F^{K}_{\alpha\beta}(\omega)   = - F^{K}_{\beta\alpha}(-\omega)
\label{eq:Fsymm}
\end{equation}

\noindent
{\it For the Majorana fermion}, due to its fundamental hermicity, $\gamma^{+}=\gamma$, only one correlator can be defined. By setting $A=B=\gamma$, we find
\begin{equation}
D_\mathsf{K}(t_{i},t_{j}') =  -i\langle\mathrm{T}_\mathsf{K}\gamma(t_{i})\gamma(t_{j}')\rangle,
\end{equation}
which combines the properties of both normal and anomalous Green functions, $\widetilde{D}_{\mathsf{K}}(t_{i},t_{j}') = \overline{D}_\mathsf{K}(t_{i},t_{j}') = D_\mathsf{K}(t_{i},t_{j}')$. Therefore, it must fulfill the same symmetry properties with respect to the frequency as the normal and the anomalous Green functions, 
\begin{align}
D^{R/A}(\omega) &= - D^{A/R}(-\omega), &
D^{K}(\omega)   &= - D^{K}(-\omega),
\label{eq:DsymmA}
\\
D^{R/A}(\omega) &=   \bigl(D^{A/R}(\omega)\bigr)^{*}, &
D^{K}(\omega)     &= - \bigl(D^{K}(\omega)\bigr)^{*}.
\label{eq:DsymmB}
\end{align}

\begin{table}
\begin{tabular}{|c|c|c|c|c|c|c|c|c|}
\hline
& 
$F^{R }_{\alpha\beta}(\omega)$ & 
$F^{A }_{\alpha\beta}(\omega)$ &
$F^{K }_{\alpha\beta}(\omega)$ & 
$F^{c }_{\alpha\beta}(\omega)$ &
$F^{ac}_{\alpha\beta}(\omega)$ & 
$F^{> }_{\alpha\beta}(\omega)$ &
$F^{< }_{\alpha\beta}(\omega)$ & 
$F^{s }_{\alpha\beta}(\omega)$ 
\\\hline
$F^{R }_{\beta\alpha}(-\omega)$ &   &$-$&   &   &   &   &   &   \\\hline
$F^{A }_{\beta\alpha}(-\omega)$ &$-$&   &   &   &   &   &   &   \\\hline
$F^{K }_{\beta\alpha}(-\omega)$ &   &   &$-$&   &   &   &   &   \\\hline
$F^{c }_{\beta\alpha}(-\omega)$ &   &   &   &$-$&   &   &   &   \\\hline
$F^{ac}_{\beta\alpha}(-\omega)$ &   &   &   &   &$-$&   &   &   \\\hline
$F^{> }_{\beta\alpha}(-\omega)$ &   &   &   &   &   &   &$-$&   \\\hline
$F^{< }_{\beta\alpha}(-\omega)$ &   &   &   &   &   &$-$&   &   \\\hline
$F^{s }_{\beta\alpha}(-\omega)$ &   &   &   &   &   &   &   &$+$ \\\hline
\end{tabular}
\caption{The transformation of the different correlators with respect to the exchange of the ladder operators.
The sign and the position of the cell denotes the connection between the Green functions.
The non-standard correlators ``ac'' (anti-causal) and ``s'' (spectral) are defined in the Eqs.~\eqref{eq:GKdef} and~\eqref{eq:Fsdef}, respectively. }
\label{tab:Fsymm}
\end{table}

Let us discuss the manifestation of the fermionic anticommutation in the anomalous Green function.
We summarized the transformations under the exchange of the particles of various anomalous correlators in Table~\ref{tab:Fsymm}.
The usual choice of the Keldysh, retarded and advanced Green functions does not fully reflect the fermionic nature of the particles.
While the Keldysh component indeed changes the sign under the exchange of particles, the retarded and advanced transform into each other.
However, we can construct two independent correlators from the symmetric and antisymmetric superposition of the retarded and advanced Green functions. From the sum of $F^{R}$ and $F^{A}$, we obtain an independent correlator which is odd with respect to particle exchange. Further, using the initial Keldysh Green function in Eq.~\eqref{eq:GKdef}, we notice that $F^R+F^A=F^{c}-F^{ac}$.

The other independent correlator is given by $F^{R}-F^{A}=F^{>}-F^{<}\equiv F^s$, and it is even with respect to particle exchange, as can be seen from
\begin{equation}
F^{s}_{\alpha\beta}(t-t') = -i\langle \psi_{\alpha}(t)\psi_{\beta}(t')+\psi_{\beta}(t')\psi_{\alpha}(t)\rangle.
\label{eq:Fsdef}
\end{equation}
$F^s$, which is expressed in terms of the ``greater'' and ``lesser'' Green functions $F^{<,>}$, describes the spectral properties of the system. 
We would like to stress that, even in the case of pure odd-frequency superconductivity, this correlator is still even in $\omega$.
This is not surprising when we consider that, for thermal equilibrium, the Keldysh Green function adopts the form $F^{K}=F^{s}\tanh(\omega/2T)$.
As long as the Keldysh component is odd in frequency, the spectral one is bound to be even.

\section{Current in SC-QD-M setup}
\label{apx:current}

The full Hamiltonian of the quantum dot on the superconducting substrate coupled to the Majorana state is $H=H_{dot}+H_{t}$, where $H_{dot}$ is given in Eq.~(11) of the main text.
The tunneling term in the most general gauge of the dot ladder operators can be written in the form 
\begin{equation}
H_{t} = \sum_{\sigma}\left(\textrm{t}_{\sigma} \gamma c_{\sigma} + \textrm{t}_{\sigma}^{*}c_{\sigma}^{\dagger}\gamma \right).
\label{eq:Hta}
\end{equation}
The general gauge transformations of the operators $c_{\sigma}$ belong to the $U(2)=U(1)\otimes SU(2)$ Lee group, which can be split into the $U(1)$ charge gauge and the $SU(2)$ spin gauge.
The Hamiltonian $H_{dot}$ is invariant under $SU(2)$ transformations, but $U(1)=\mathrm{e}^{i\varphi}$ changes the superconducting phase by $\phi\to\phi+2\varphi$, as expected.
The tunneling Hamiltonian $H_{t}$, due to the hermicity of the Majorana operator $\gamma$, is not invariant under either $U(1)$ or $SU(2)$ transformations. As a result, we have the freedom to change the tunneling coefficients $\textrm{t}_{\sigma}$ by choosing the appropriate spin gauge. 
In an experimental realization of a MBS, this $SU(2)$ symmetry is usually broken not by the tunneling amplitude, but by the magnetic order and the spin-orbit interaction in the STM tip, a typical setup for the creation of the MBS~\cite{Lutchyn2010,Oreg2010}.
Nevertheless, in the effective model of Eq.~\eqref{eq:Hta}, any tunneling coefficients $\textrm{t}_{\uparrow}$ and  $\textrm{t}_{\downarrow}$ can be transformed into
\begin{equation}
\sum_{\sigma'}U_{\sigma\sigma'}\textrm{t}_{\sigma'} =
\begin{pmatrix}
\textrm{t} \\ 0
\end{pmatrix},
\quad\text{where}\quad
U = \frac{i}{\textrm{t}}
\begin{pmatrix}
\textrm{t}_{\uparrow}^{*} & \textrm{t}_{\downarrow}^{*} \\
\textrm{t}_{\downarrow} & -\textrm{t}_{\uparrow}
\end{pmatrix}
\in SU(2)
\quad\text{and}\quad
\textrm{t}=\sqrt{|\textrm{t}_{\uparrow}|^{2}+|\textrm{t}_{\downarrow}|^{2}}\in \Re>0.
\end{equation}
Let us demonstrate how the gauge transformations work in the Fock subspace with odd fermion parity defined in the main text, and how one obtains the Hamiltonian in the form of Eq.~(15).

The total Hamiltonian, in the arbitrary gauge, is
\begin{equation}
H=\bordermatrix{
 & |\uparrow\downarrow,1\rangle & |\uparrow,0\rangle & |\downarrow,0\rangle & |0,1\rangle \cr
\langle\uparrow\downarrow,1| & \delta & -\textrm{t}_{\downarrow}^{*} & \textrm{t}_{\uparrow}^{*} & -\frac{\Gamma_{\Delta}}{2}e^{i\phi} \cr
\langle\uparrow,0| & -\textrm{t}_{\downarrow} & \varepsilon+\frac{B_{z}'}{2} &  \frac{B_{\perp}'}{2}e^{-i\theta'} & -\textrm{t}_{\uparrow}^{*} \cr
\langle\downarrow,0| & \textrm{t}_{\uparrow} &  \frac{B_{\perp}'}{2}e^{i\theta'} & \varepsilon-\frac{B_{z}'}{2} & -\textrm{t}_{\downarrow}^{*} \cr
\langle0,1| & -\frac{\Gamma_{\Delta}}{2}e^{-i\phi} & -\textrm{t}_{\uparrow} & -\textrm{t}_{\downarrow} & 0
}
\end{equation}
Using the block-diagonal matrix $V_{U}=\mathrm{diag}(1,U,1)$ corresponding to the $SU(2)$ rotation in the second quantization, we can eliminate $\textrm{t}_{\downarrow}$ and set $\textrm{t}_{\uparrow}$ to $\textrm{t}$, which is real and positive, resulting in
\begin{equation}
V_{U}^{+}HV_{U}=
\begin{pmatrix}
\delta & 0 & \textrm{t} & -\frac{\Gamma_{\Delta}}{2}\mathrm{e}^{i\phi} \\
0 & \varepsilon+\frac{B_{z}}{2} & \frac{B_{\perp}}{2}\mathrm{e}^{-i\theta} & -\textrm{t} \\
\textrm{t} & \frac{B_{\perp}}{2}\mathrm{e}^{i\theta} & \varepsilon-\frac{B_{z}}{2} & 0 \\
-\frac{\Gamma_{\Delta}}{2}\mathrm{e}^{-i\phi} & -\textrm{t} & 0 & 0
\end{pmatrix}
=
V_{\theta}
\begin{pmatrix}
\delta & 0 & \textrm{t} & -\frac{\Gamma_{\Delta}}{2}\mathrm{e}^{i\phi-i\theta} \\
0 & \varepsilon+\frac{B_{z}}{2} & \frac{B_{\perp}}{2} & -\textrm{t} \\
\textrm{t} & \frac{B_{\perp}}{2} & \varepsilon-\frac{B_{z}}{2} & 0 \\
-\frac{\Gamma_{\Delta}}{2}\mathrm{e}^{-i\phi+i\theta} & -\textrm{t} & 0 & 0
\end{pmatrix}
V_{\theta}^{+}.
\end{equation}
In the second step, we have used the matrix $V_{\theta}=\mathrm{diag}(\mathrm{e}^{i\theta},1,\mathrm{e}^{i\theta},1)$, which is a combination of the rotation by an angle $\theta$ around the $z$-axis [SU(2) by $\mathrm{diag}(1,\mathrm{e}^{-i\theta/2},\mathrm{e}^{i\theta/2},1)$], and a charge gauge change by $\theta/2$ [U(1) by $\mathrm{diag}(\mathrm{e}^{i\theta},\mathrm{e}^{i\theta/2},\mathrm{e}^{i\theta/2},1)$].

We now provide the details for the derivation of the expression for the current in Eq.~(16) of the main text.
We define the current operator as
\begin{equation}
\hat{I} = i\frac{e}{\hbar} \sum_{\sigma}\left(\textrm{t}_{\sigma}\gamma c_{\sigma} - \textrm{t}_{\sigma}^{*}c_{\sigma}^{\dagger}\gamma\right)
=
i\frac{e}{\hbar}
\begin{pmatrix}
 0 & \textrm{t}_{\downarrow}^{*} & -\textrm{t}_{\uparrow}^{*} & 0 \\
 -\textrm{t}_{\downarrow} & 0 & 0 & \textrm{t}_{\uparrow}^{*} \\
 \textrm{t}_{\uparrow} & 0 & 0 & \textrm{t}_{\downarrow}^{*} \\
 0 & -\textrm{t}_{\uparrow} & -\textrm{t}_{\downarrow} & 0
\end{pmatrix} .
\end{equation}
The superconducting phase dependence can be moved from the order parameter to the tunneling coefficients using the $U(1)$ charge gauge transformation $V_{\phi}=\mathrm{diag}(\mathrm{e}^{i\phi},\mathrm{e}^{i\phi/2},\mathrm{e}^{i\phi/2},1)$.  As a result, we find
\begin{equation}
V_{\phi}^{+}HV_{\phi}=
\begin{pmatrix}
\delta & -\textrm{t}_{\downarrow}\mathrm{e}^{-i\phi/2} & \textrm{t}_{\uparrow}\mathrm{e}^{-i\phi/2} & -\frac{\Gamma_{\Delta}}{2} \\
-\textrm{t}_{\downarrow}e^{i\phi/2} & \varepsilon+\frac{B_{z}}{2} &  \frac{B_{\perp}}{2}\mathrm{e}^{-i\theta} & -\textrm{t}_{\uparrow}\mathrm{e}^{-i\phi/2} \\
\textrm{t}_{\uparrow}e^{i\phi/2} &  \frac{B_{\perp}}{2}\mathrm{e}^{i\theta} & \varepsilon-\frac{B_{z}}{2} & -\textrm{t}_{\downarrow}\mathrm{e}^{-i\phi/2} \\
-\frac{\Gamma_{\Delta}}{2} & -\textrm{t}_{\uparrow}\mathrm{e}^{i\phi/2} & -\textrm{t}_{\downarrow}\mathrm{e}^{i\phi/2} & 0
\end{pmatrix} ,
\end{equation}
which provides the relation $2\frac{e}{\hbar}\partial_{\phi} (V_{\phi}^{+}HV_{\phi}) = V_{\phi}^{+}\hat{I}V_{\phi}$.
Starting from the equation for the current defined in the main text, we arrive at
\begin{equation}
I=
-2\frac{e}{\hbar}\beta^{-1}\partial_{\phi} \log Z = 
-2\frac{e}{\hbar}\beta^{-1}Z^{-1}\partial_{\phi} \mathrm{Tr\,} \mathrm{e}^{-\beta H} = 
-2\frac{e}{\hbar}\beta^{-1}Z^{-1} \mathrm{Tr\,} \partial_{\phi} \mathrm{e}^{-\beta V_{\phi}^{+}HV_{\phi}} ,
\end{equation}
with $Z=\sum_{\alpha}\mathrm{e}^{-E_{\alpha}/T}$. Using the standard formula for the derivative of the exponential map~\cite{Wilcox1967}, $\partial \mathrm{e}^{X} = \int_{0}^{1} \mathrm{e}^{sX} (\partial X) \mathrm{e}^{(1-s)X}ds$, and the relation between current and Hamiltonian stated above, we get
\begin{equation}
I=
Z^{-1} \mathrm{Tr\,} \int_{0}^{1} \mathrm{e}^{-s\beta V_{\phi}^{+}HV_{\phi}} V_{\phi}^{+}\hat{I}V_{\phi} \mathrm{e}^{-(1-s)\beta V_{\phi}^{+}HV_{\phi}}ds=
Z^{-1} \mathrm{Tr\,} V_{\phi}^{+}\hat{I}V_{\phi} \mathrm{e}^{-\beta V_{\phi}^{+}HV_{\phi}} = 
Z^{-1} \mathrm{Tr\,} \hat{I} \mathrm{e}^{-\beta H}.
\end{equation}


\end{document}